\documentclass[aps,eqsecnum,nofootinbib,superscriptaddress,showpacs]{revtex4}
\def\be{\begin{eqnarray}}
\def\ee{\end{eqnarray}}

\def\p{\partial}

\usepackage{amsmath,amsfonts,amssymb,bm}
\usepackage{wrapfig}
\usepackage{graphicx}
\usepackage[dvips]{color}
\begin{document}
\title{Singularities in asymptotically anti-de Sitter spacetimes
}

\date{\today}

\begin{abstract}
We consider singularity theorems in asymptotically anti-de Sitter (AdS) 
spacetimes. In the first part, we discuss the global methods used to 
show geodesic incompleteness and see that when the conditions 
imposed in Hawking and Penrose's singularity theorem are satisfied, 
a singularity must appear in asymptotically AdS spacetime. 
The recent observations of turbulent instability of asymptotically 
AdS spacetimes indicate that AdS spacetimes are generically singular 
even if a closed trapped surface, which is one of the main conditions 
of the Hawking and Penrose theorem, does not exist 
in the initial hypersurface. This may lead one to expect to obtain 
a singularity theorem without imposing the existence of a trapped set 
in asymptotically AdS spacetimes. This, however, does not appear to 
be the case. 
We consider, within the use of global methods, two such attempts and 
discuss difficulties in eliminating conditions concerning a trapped set 
from singularity theorems in asymptotically AdS spacetimes.   
Then in the second part, we restrict our attention to the specific case of 
spherically symmetric, perfect fluid systems in asymptotically AdS spacetime, 
and show that under a certain condition concerning dynamics of the fluid, 
a closed trapped surface must form, and as a combined result with 
Hawking and Penrose's theorem, that such a spacetime must be singular. 
\end{abstract}

\author{Akihiro Ishibashi}
\email{akihiro@phys.kindai.ac.jp}
\affiliation{Department of Physics, Kinki University, Higashi-Osaka, 
577-8502, Japan}

\author{Kengo Maeda} 
\email{maeda302@sic.shibaura-it.ac.jp}
\affiliation{Faculty of Engineering,
Shibaura Institute of Technology, Saitama, 330-8570, Japan}

\maketitle

\section{Introduction}
\begin{figure}[h]\begin{center} 
\includegraphics[width=2.5cm]{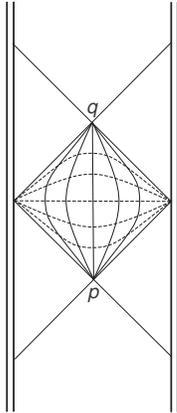} \hspace{1cm} 
\caption{\label{FLRW-chart}
\small{The AdS-infinity is represented by a timelike
                  surface. The diamond region is covered by an open FLRW
                  chart. Every timelike geodesic through $p$ intersects
                  all the timeslices defined by this chart and reach
                  $q$, which corresponds to the anti-podal point to
                  $p$. 
                  }}
\end{center}\end{figure}

Anti-de Sitter (AdS) spacetimes play a role of theoretical laboratory, 
where different areas of physics appear to make close contact, 
as exemplified in the AdS/CFT correspondence and its various 
applications~\cite{adscft}.  
In this context, it is essential, besides its asymptotic symmetries, that 
the conformal infinity of AdS spacetime is timelike and therefore 
can be viewed as another---lower dimensional---spacetime where
non-gravitational theories dual to gravity theories in AdS are supposed 
to reside. AdS spacetimes are maximally symmetric, negatively curved, 
geodesically complete but non-globally hyperbolic, due to this timelike 
nature of infinity. 
As such, for any given point $p$ of AdS spacetime, 
all {\em timelike geodesics} from $p$ never reach the infinity and 
necessarily intersect the antipodal point $q$ to $p$
\footnote{
Hereafter, by a $D$-dimensional AdS spacetime we mean 
the covering space of anti-de Sitter spacetime so that its topology 
is ${\Bbb R}^{D}$, rather than $S^1 \times {\Bbb R}^{D-1}$, 
and therefore contains no closed timelike curves. 
}
. This can easily be seen by considering the cosmological chart 
\begin{align}  
  ds^2 = -d\tau^2 
       + \cos^2 \tau \left( d\chi^2 + \sinh^2 \chi d\Omega_{(D-2)}^2 \right) 
\,,  
\end{align}  
where for simplicity the curvature radius is normalized to unity. 
This describes an open FLRW universe with big-bang and big-crunch 
respectively at $\tau \rightarrow \pm \pi /2$, where the scale factor 
vanishes. 
In fact, because of its maximum isometries, for any given point $p$ in AdS, 
one can find the same cosmological chart in which $p$ corresponds to 
the big-bang or crunch point [see figure~\ref{FLRW-chart}]. 
These big-bang/crunch points are merely a singular point 
of congruences of geodesic curves with the tangent 
$(\partial /\partial \tau)^a$ or a caustic thereof, but not a singularity of 
the spacetime structure itself. However, since all the comoving lines 
converge there, one can anticipate that if perturbed by, e.g., adding some 
matter fields, then the perturbation fields may grow along the comoving lines 
and eventually make big-bang/crunch points true curvature singularities.  
%
If this is the case it may be viewed also as a Cauchy horizon instability 
since the boundary of the causal past of $q$ is the Cauchy horizon 
for any $\tau=const.$ partial Cauchy surface.   


\medskip 
To study such dynamics of perturbations in AdS spacetimes, suitable boundary 
conditions on perturbation fields need to be imposed at AdS infinity.  
It has been shown that under the Dirichlet (and some other types of) 
boundary conditions, AdS spacetime is stable at least with respect to 
linear perturbations~\cite{IW04}. 
However, there has recently appeared 
evidence that asymptotically AdS spacetimes are {\em non-linearly} 
unstable~\cite{Bizon,Dias}.  
In particular, it has been shown by Bizon and Rostworowski~\cite{Bizon} that 
upon certain assumptions, a black hole---hence singularities as well---must 
be formed in asymptotically AdS spaces: They considered a spherically 
symmetric gravitating system of a minimally coupled massless scalar field 
and numerically shown that any Gaussian-type wave packet of the field does 
necessarily collapse to form a black hole. 

\medskip 
This should be compared to the asymptotically flat case, in which 
it has been shown that small but non-linear perturbations remain small 
in its amplitude, and gravitational collapse of matter fields does not 
necessarily result in the black hole formation. 
More specifically, according to Choptuik's numerical 
calculation~\cite{choptuik} of a scalar field collapse, 
the initial data set can be divided into two classes:  
supercritical and subcritical data. 
In the former case where the Gaussian-type wave packet 
has large amplitude, the scalar field collapses to form a black hole, whereas 
in the latter case, having small amplitude, 
the Gaussian wave packet bounces off the center of the spherical symmetry 
and escapes to infinity without forming a black hole. 
In contrast, the results of AdS nonlinear instability above \cite{Bizon} 
indicates that asymptotically AdS spacetimes are generically singular. 
One is therefore led to anticipate type of singularity theorems 
to be established in a class of asymptotically AdS spacetimes 
without there being any region of strong gravity to trap on initial data 
surface.  

\medskip 
The purpose of this paper is to discuss singularity theorems in 
asymptotically AdS spacetimes. 
In the first part, we will attempt to clarify which results of the well-known 
singularity theorems can apply to the asymptotically AdS case.   
We briefly review the basics of singularity theorems and see that 
Hawking and Penrose's version of the singularity theorems applies to 
the AdS case without any major change in its assumptions, in 
section~\ref{sec:basics}. The singularity theorems typically assume 
the existence of a strong gravity region, 
such as a closed trapped surface. However, the results of \cite{Bizon} 
indicates that a black hole forms even starting from arbitrarily small 
initial perturbations, and thus one may hope to obtain a singularity theorem 
with {\em weaker}, or even {\em no}, assumptions concerning strong gravity region.  
We discuss what the main obstacle is when attempting to eliminate 
the assumption of trapped sets from singularity theorems 
in asymptotically AdS case, in section~\ref{subsec:diff}. 
In the second part, we will focus on a specific case: a system of spherically 
symmetric spacetime with a perfect fluid, and present a singularity theorem, 
in section~\ref{sec:sss}. 
%
To show our singularity theorem we have to impose a condition that 
guarantees that the spacetime to consider must be dynamical in a certain 
sense. Also we will see that our proof need some restriction on 
the equation of state of the perfect fluid. 
We summarize our results in section~\ref{sec:conclusion} and discuss 
future directions of the present subject. In Appendix, we discuss whether 
one can find a regular, static configuration in asymptotically AdS spacetime 
in a similar setting discussed in the main part of the paper.

\section{Asymptotically Anti-de Sitter spaces and singularities} 

\subsection{Hawking and Penrose's theorems and AdS spacetimes} 
\label{sec:basics}
We start with briefly recapitulating basic ideas of 
the singularity theorems and then discuss whether (or which of) 
the singularity theorems can fit the case of asymptotically AdS spacetimes. 
A spacetime singularity is defined by, as one of its essential features, 
the causal geodesic incompleteness. 
The singularity theorems show that under physically reasonable conditions, 
the spacetime must be (either timelike or null, or both) 
geodesically incomplete. 
As ``physically reasonable'' conditions, typically imposed are 
\begin{itemize}
\item[(a)] The energy conditions
\item[(b)] The conditions on global structure, 
\item[(c)] The existence of a strong gravity region 
\end{itemize}
The condition~(a) describes, via the Einstein equations, the convergence of 
causal geodesic congruences. For some versions of the singularity theorems, 
the generic conditions may also be required. 
The condition~(b) is the requirement of either 
the strong causality or chronology condition or 
the existence of a Cauchy surface. The condition~(c) requires a closed trapped 
surface for gravitational collapse, and in the cosmological context, 
a compact achronal hypersurface without edge or a point for which a null 
geodesic congruence starts to reconverge.

\medskip 
To show the geodesic incompleteness, the following two notions play 
a central role: 
\begin{itemize}
\item[{(i)}] Existence of (a pair of) conjugate points,   
\item[{(ii)}] Global hyperbolicity.  
\end{itemize} 
The occurrence of a pair of conjugate points concerns the behavior of 
a geodesic congruence, which is governed by the Raychauduri equation: 
\begin{align}
\label{eqn:Raychaudhuri}
\frac{d\theta}{d s}= - R_{\mu \nu}K^\mu K^\nu - 2\sigma^2 
                          -\frac{1}{n}\theta^2 \,,   
\end{align}
where $s$ denotes an affine parameter of a geodesic curve with tangent 
$K^\mu$, $\theta$ the expansion of the geodesic congruence, 
$\sigma$ the shear, and the number $n$ depends on if $K^\mu$ is timelike or 
null and also on the number of spacetime dimensions $D$.  
At conjugate points, the expansion diverges: $\theta \rightarrow \pm \infty$. 
It is ensured by the condition (a) above---together with the generic 
condition, and is essentially due to the fact that gravity is attractive. 
The existence of conjugate points leads the following important result. 
Consider a future directed causal curve, $\gamma(t)$, from $p$ to 
$q \in J^+(p)$, whose length is given by 
\begin{align}
\label{def:length}
  L(\gamma: p \rightarrow q) = \int^q_p \sqrt{
     -g\left(
            \frac{\partial}{\partial t},\frac{\partial}{\partial t}  
      \right)}
     d t \,.    
\end{align}
Now suppose that $\gamma$ is geodesic and contains a point $r$ conjugate 
to $p$ between $p$ and $q$. 
Then one can find a future directed causal curve from $p$ to $q$ 
whose length is longer than $\gamma$, by considering the second variation of 
the length $L(\gamma: p \rightarrow q)$.  
A similar result holds between a point and a hypersurface. 

\medskip  
A set $\cal N$ in $(M,g)$ is said to be globally hyperbolic if the strong 
causality holds on $\cal N$ and for any two points $p,q \in {\cal N}$, 
$J^+(p) \cap J^-(q)$ is compact, contained in $\cal N$. 
The importance of this notion for the singularity theorems is that it enables 
one to prove existence of maximal length causal curves. 
Namely, for any causally related two points, $p$ and $q \in J^+(p)$, 
in a globally hyperbolic region ${\cal N}$, there exists a geodesic 
curve from $p$ to $q$ in $\cal N$ which attains the maximal length 
among all continuous causal curves from $p$ to $q$ in ${\cal N}$.  
Again a similar result holds between a point and a hypersurface. 
This, combined together with the result from (i), implies that in a globally 
hyperbolic region $\cal N$, no causal geodesic curve connecting 
two points $p,q \in {\cal N}$ can contain a pair of conjugate points 
within the two points if it has the maximal length among all continuous 
causal curves from $p$ to $q$. 
The singularity theorems show that under suitable versions of 
the conditions~(a), (b), and (c), 
if all geodesics are complete, then there must be a globally hyperbolic region 
within which every causal geodesic curve has to admit a pair of conjugate 
points on it, thus establishing a contradiction that falsifies 
the assumption of the geodesic completeness. 


\medskip 
There are several different versions of the singularity theorems 
depending on different combinations of the conditions~(a), (b), and (c). 
The first theorem [see Theorem~1 in chapter 8 of \cite{HE}], proposed 
by Penrose for gravitational collapse, requires the null convergence 
for the condition~(a), the existence of a closed trapped surface as for (c), 
and the existence of non-compact Cauchy surface as for (b). 
The last one in particular requires that the entire spacetime $(M,g)$ 
be globally hyperbolic, which is clearly not the case for 
asymptotically anti-de Sitter spacetimes.

\medskip 
Hawking's theorem for cosmological singularities [see Theorem 4 in 
chapter 8 of \cite{HE}] removes the assumption of global hyperbolicity by, 
instead of requiring an expanding Cauchy surface, 
an expanding compact spacelike hypersurface without edge. 
AdS conformal boundary acts just like a boundary of a box that confines bulk 
fields. For this reason one might anticipate that Hawking's version of 
the singularity theorem may be generalized to apply to asymptotically AdS 
spacetimes. 
This theorem again does not 
straightforwardly apply as the AdS is not spatially compact\footnote{ 
Note that if the magnitude of a negative cosmological constant $\Lambda$ does 
not dominate over other cosmic matter contents on some initial data surface, 
one might anticipate that either Penrose's or Hawking's theorem 
just mentioned above would apply since $\Lambda <0$ generally enhances 
the convergence of timelike geodesics and does not affect the convergence 
of null geodesics [as seen in (\ref{eqn:Raychaudhuri}) combined together 
with (\ref{condi:timelike:convergence}) below]. 
In fact, such a generalization of singularity theorems 
for open universes with a Cauchy surface or closed universes 
has been considered in the earlier work~\cite{Tipler76}, 
in which the effects of a negative cosmological constant on the timelike 
convergence were taken into account.  
In that case, however, the requirement of either a Cauchy surface/global 
hyperbolicity or a compact surface without edge would not allow 
timelike conformal infinity to be attached, and therefore 
the entire spacetime would fail to be an asymptotically AdS 
spacetime considered in the present context.   
}. 


\medskip 
A version which has much wider applicability is the one proposed by 
Hawking and Penrose [see Theorem~2 in chapter 8 of \cite{HE}], in which 
the condition~(b) requires, instead of global hyperbolicity, the chronology 
condition and the condition~(c) can be re-phrased as the existence of 
a future (or past) trapped set $\cal S$, an achronal set for which 
the horismos $E^+({\cal S}):= J^+({\cal S}) - I^+({\cal S})$ 
(or $E^-({\cal S})$) is compact. 
Then, under the convergence and generic conditions to null geodesics, 
one can construct a globally hyperbolic region ${\cal N}$ which contains 
a future as well as past inextendible causal curve.

\medskip 
Now we argue that this version of the singularity theorems, which requires 
the existence of a closed trapped surface as a trapped set ${\cal S}$ required 
in (c), applies to the asymptotically AdS case without any major change 
in its assumptions, as far as the stress-energy tensor for matter fields 
together with the negative cosmological constant, does not violate 
the null, as well as the timelike convergence property.   
The point is that in asymptotically anti-de Sitter spacetimes 
the affine distance from any point in the spacetime to AdS boundary 
along a null geodesic is infinite, even though the coordinate distance 
may be finite. Therefore any outgoing null geodesics orthogonally emanating 
from the trapped surface, ${\cal S}$, must have a conjugate point 
with respect to ${\cal S}$ before reaching AdS-boundary, 
making the future horismos, $E^+({\cal S})$, compact. 
Then, there is no obstruction to apply (Corollary of) Lemma~8.2.1 of \cite{HE} 
to show the existence of a future inextendible timelike curve $\gamma$ 
contained in the globally hyperbolic region $D^+(E^+({\cal S}))$. 
Then, again using the fact that AdS-boundary is at affinely infinite distances 
along null geodesics from any point, 
we see that the past-inextendible extension of any null generating segment 
of ${\dot J}^-(\gamma)$ must enter $I^-(\gamma)$ and have a past end point, 
thus making ${\cal F}:=E^+({\cal S})\cap {\bar J}^-(\gamma)$ past-trapped: 
$E^-({\cal F})$ is compact.  
Then, again applying Lemma~8.2.1 of \cite{HE}, we also have a past 
inextendible timelike curve $\lambda$ in $D^-(E^-(\cal F))$. 
\begin{figure}[h]\begin{center}
\includegraphics[width=9cm]{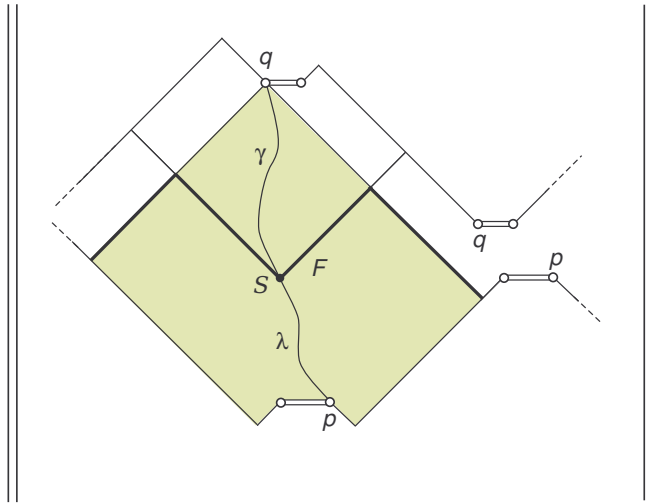} \hspace{1cm} 
\caption{\label{GHregion}
\small{ ${\cal N}= D(E^-({\cal F}))$ (Green color): 
To find this globally hyperbolic subregion in $M$, we have used 
the fact that in asymptotically anti-de Sitter spacetimes 
the affine distance from any point in the spacetime to AdS boundary 
along a null geodesic is infinite, 
as well as that the convergence property of null geodesics is not affected 
by the presence of the negative cosmological constant. Within this globally hyperbolic 
region, one can apply the argument about the maximum length curve. }}
\end{center}\end{figure}
Note that to find the desired globally hyperbolic region 
${\cal N}= D(E^-({\cal F}))$ we have used only the fact that 
in asymptotically anti-de Sitter spacetimes 
the affine distance from any point in the spacetime to AdS boundary 
along a null geodesic is infinite and 
that the convergence property of null geodesics is not affected by 
the presence of the negative cosmological constant.  
The rest of arguments concerning the maximum length non-spacelike curves 
in the globally hyperbolic region ${\rm int} ( D(E^-(\cal F)))$ 
parallels the proof in Hawking and Ellis~\cite{HE}, 
but for later discussion, we shall sketch the arguments.  
Consider a sequence of points $x_n$ on $\lambda$, $x_{n+1} \in I^-(x_n)$, 
so that $\{x_n\}$ tends to the past as $n\rightarrow \infty$, 
and similarly a sequence of points $\{y_n\}$ on $\gamma$ tending to the 
future, $y_{n+1} \in I^+(y_n)$. 
Then as $x_n, \; y_n$ are in ${\cal N}$, one can find a timelike geodesic,  
$\mu_n$, from $x_n$ to $y_n$, which maximizes the length from $x_n$ to $y_n$. 
All $\mu_n$ cross the compact set ${\cal F}$, and therefore 
there exists a limit curve $\mu$ of $\{\mu_n\}$, which must also be a
 timelike geodesic contained within $\cal N$. 
Now if $\mu$ is incomplete, then we are done. If $\mu$ is complete, 
then $\mu$ must contain a pair of conjugate points due to the convergent 
condition (a) plus the timelike generic condition. 
This, however, implies that a timelike geodesic $\mu_n$ with sufficiently 
large $n$ would also have to admit a pair of conjugate points as 
conjugate points vary continuously. This is a contradiction 
against that $\mu_n$ was supposed to be the timelike curve 
of maximum length from $x_n$ to $y_n$.

\medskip 
Finally, we comment that the timelike convergence conditions assumed 
in the above arguments hold if, for any timelike vector $K^\mu$, 
the stress-energy tensor $T_{\mu \nu}$ for matter fields satisfies  
\begin{align}
T_{\mu \nu}K^\mu K^\nu \geq \frac{1}{D-2}\left( 
                                    T - \frac{1}{4\pi} \Lambda 
                              \right) K_\mu K^\mu  \,.
\label{condi:timelike:convergence}
\end{align}
In fact, this is less restrictive than requiring $T_{\mu \nu}$ alone 
to satisfy the standard strong energy condition, due to the last term. 
In other words, the presence of $\Lambda<0$ 
enhances the timelike convergence $R_{\mu \nu}K^\mu K^\nu \geq 0$ 
via the Einstein equations.  

\medskip 

Thus we have seen the following version of the singularity theorem
applies to asymptotically AdS spacetimes: 

\medskip
\noindent 
{\bf Theorem.} [Hawking and Penrose:] \\ 
{\it 
A spacetime $(M,g_{\mu \nu})$ is not timelike and null 
geodesically complete if the following conditions hold: 
\begin{itemize}
\item[(a)] $R_{\mu \nu}K^\mu K^\nu \geq 0$ for every non-spacelike vector $K^\mu$, and the generic condition holds for every non-spacelike geodesic, i.e.,
every non-spacelike geodesic has at least one point where $K_{[\lambda} R_{ \nu ] \alpha \beta [\mu}
           K_{\sigma]}K^\alpha K^\beta \neq 0$. 

\item[(b)] The chronology condition holds.

\item[(c)] There exists a closed trapped surface.
\end{itemize} 
} 

\medskip 
\noindent 
Note that the condition~(c) can be replaced with the other case: 
there exists a point for which light cones start reconverging.  

\subsection{Difficulties in weakening the strong gravity conditions} 
\label{subsec:diff}

\subsubsection{Non-global hyperbolicity and maximum length of causal curves}  


The non-linear instability of AdS spacetime from arbitrarily small initial 
perturbations \cite{Bizon,Dias} indicates that singularities may generically 
form without there being any strong gravity region on initial data surface 
and thus motivates us to consider if one can obtain a singularity theorem 
without the condition for strong gravity. 
In this subsection we will discuss some difficulties in eliminating 
the condition~(c).

\medskip 
The key point of the black hole formation from subcritical initial 
data in~\cite{Bizon} is that the collapsing scalar field has to experience 
the process of bouncing-off at AdS infinity many times 
to make its configuration supercritical. 
This is not possible within a single, globally hyperbolic 
region such as the diamond region of the cosmological chart in 
figure~\ref{FLRW-chart}. Therefore, in order to obtain a singularity 
theorem without condition~(c), one has to deal with a non-globally 
hyperbolic (sub)region. But then, the lack of global hyperbolicity 
%
%
%
falsifies standard arguments about the existence of maximum length 
geodesic curve in relevant region, which is one of the central 
ingredients of Hawking and Penrose's proof discussed in the previous section. 

\medskip 
Let us consider this problem in a concrete example of pure AdS spacetime 
$(M,g)$. The AdS metric can be expressed by the standard global chart,  
$(t,\chi,\Omega)$, in the static form 
\begin{align}
 ds^2 
 = \frac{1}{\cos \chi^2} \left(-dt^2 + d\chi^2 + \sin^2 \chi d\Omega^2 \right) 
\,. 
\end{align}
This is conformally embedded into the Einstein static universe 
$(\tilde M, {\tilde g})$, so that the conformal boundary $\partial M$ 
is attached to $M$ as $\tilde M= M \cup \partial M$, and 
located at $\chi = \pm \pi/2$ in $\tilde M$.

\begin{figure}[h]\begin{center}
\includegraphics[width=2.5cm]{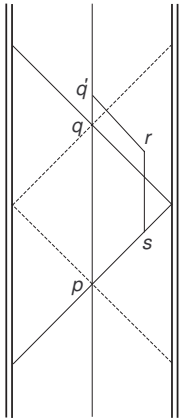} \hspace{1cm} 
\caption{\label{Length}
\small{In AdS spacetime, one can construct a causal curve from $p$ to $q'$ which is not geodesic and has an arbitrarily large length.}}
\end{center}\end{figure}

\medskip 
Now consider two points $p$ and $q$ which respectively correspond to 
the big-bang/crunch point in the open FLRW chart so that $p$ and $q$ form 
a pair of conjugate points. There are infinitely many timelike curves 
which maximize the length from $p$ to $q$, in fact all timelike geodesics 
emanating from $p$ reach the point $q$.
Therefore 
no causal curves on $M$ contained in $J^+(p) \cap J^-(q)$ reaches 
the AdS conformal boundary. 
However, as mentioned just above, the black hole formation is not realized in 
a region which correspond to a single globally hyperbolic patch.  
Therefore, we are concerned with a pair of points $p$ and $q'$, 
as depicted in Figure~\ref{Length}, for which $J^+(p) \cap J^-(q')$ 
contains a portion of $\partial M$. 
What we need to deal with is the causal curve that would maximize all causal 
curves connecting $p$ and $q'$. It turns out that the timelike geodesic from 
$p$ to $q'$ along the line $\chi=0$ is no longer the one.
To see this, let us take a point $s$ on a null geodesic curve from $p$
to
AdS infinity. Let $\chi_s$ be the coordinate value of $\chi$ at the point $s$. 
Then, move along $\chi_s=const.$ line, which is not a geodesic, until 
the point $r$ as in Figure~\ref{Length}, and move back to the center along 
a null geodesic to $q'$. 
Thus we obtain a causal non-geodesic curve $p \rightarrow s \rightarrow r \rightarrow q'$. 
One can easily evaluate the length of this path and show that this causal 
curve provides a longer length from $p$ to $q'$ than the timelike geodesic 
along $ \chi=0$. Since the null segment of this curve, those from $p$ 
to $s$ and $r$ to $q'$, does not add any length to the curve, we only focus 
on the timelike segment of the $\chi_s$ line from $s$ to $r$. 
Then, since the length is given by eq.~(\ref{def:length}), we have 
\begin{align}
 L (s \rightarrow r) \propto \frac{1}{\cos \chi_s} \,, 
\end{align} 
which can become arbitrarily large as $\chi_s \rightarrow \pi/2$.
Thus, by taking the timelike segment $s \rightarrow r$ to be close to AdS
conformal boundary, the length of the causal curve
from $p \rightarrow r \rightarrow s \rightarrow q'$
can be made arbitrarily large.
This implies that there does not exist a causal curve that maximizes
the length of all continuous causal curves from $p$ to $q'$. 
Therefore even if one can show existence of inextendible causal curve
in some region of AdS spacetime, one cannot apply the argument of
a sequence of timelike geodesics $\mu_n$ to lead a contradiction.  

\medskip
Even in non-globally hyperbolic region, it might be possible to show
the existence of maximum length curve between $p$ and $q'$, if, for
instance, one introduces a cut off radius $\chi_c$ so that any causal curves 
are not allowed to approach AdS conformal boundary beyond the cut-off 
coordinate value $\chi_c$. 
In that case, however, the maximum length curve would fail to be geodesic, 
and we would not be able to apply general results concerning variation of 
arc-length, established for geodesic curves in chapter 4 of \cite{HE}.

\subsubsection{Double covering approach}

One way to avoid the issues that arise from the non-global hyperbolicity 
discussed above would be to consider a singularity theorem in some unphysical, 
globally hyperbolic spacetime, rather than in the physical asymptotically 
AdS spacetime itself. 
Namely, one attempts to 
\begin{itemize}
\item[{(1)}] construct a spatially compact unphysical spacetime 
from asymptotically AdS spacetime in a way that 
(i) the timelike convergence and (ii) the generic conditions 
are satisfied in the unphysical spacetime, and 
\item[{(2)}] establish the relation between geodesic incompleteness 
in the unphysical spacetime and that in the physical asymptotically AdS 
spacetime. 
\end{itemize} 

\medskip 
In fact, for a given asymptotically AdS spacetime $M$, one
can first make a suitable conformal transformation to attach the
boundary $\partial M$ to have ${\tilde M} = M \cup \partial M$, and then
make double $\tilde M$ and glue them together at the AdS conformal
boundaries $\partial M$. By doing so one can obtain a single connected 
spacetime ${\widehat M}$ with closed spatial section, 
just like the Einstein static cylinder, to which one may be able to apply 
type of the singularity theorems for closed universe. 
Since $\widehat M$ is spatially compact, any point or any subset 
${\cal S} \subset \widehat M$ would be a trapped set, 
as $E^\pm ({\cal S})$ is compact in ${\widehat M}$.
Then, it is straightforward to show that under condition (a), 
the Cauchy horizon for $E^\pm(\cal S)$ must be non-compact or empty, 
and find a globally hyperbolic region ${\widehat {\cal N}}= D(E^-({\cal F}))$ in $\widehat M$ that contains future and past inextendible timelike curve, 
say, $\gamma$. Then, one may hope to apply the argument of maximum length 
curve of timelike geodesics $\mu_n$ to show the geodesic incompleteness 
of $\widehat M$.


%
\medskip 
However, the following would be a potential obstruction in this approach. 
The double cover is taken from an unphysical spacetime
whose metric is conformally isometric to the original, physical
metric. Suppose one can find a suitable conformal transformation so that
the timelike convergence conditions (\ref{condi:timelike:convergence})
are satisfied 
in unphysical spacetime. A subtle issue arises 
in the neighborhood of the AdS infinity $\partial M$ in $\widehat M$.
Since we impose the asymptotic AdS boundary conditions
at $\partial M$, we can find a neighborhood of $\partial M$ which is
isometric to a neighborhood of the $\chi= \pi/2$ line of
the Einstein static universe. Then, it is straightforward to check
that even on the neighborhood of the $\chi= \pi/2$ line,
the timelike convergence condition (\ref{condi:timelike:convergence})
is satisfied.
However, there, the condition (\ref{condi:timelike:convergence}) is
only marginally satisfied, and the generic condition is actually not.  
Hence the timelike geodesics that lie on the AdS boundary, i.e.,
 those correspond to $\chi = \pi/2$ lines, do not contain a pair of
 conjugate points. 
Therefore, even when one finds in ${\widehat {\cal N}}= D(E^-({\cal F})) \subseteq \widehat M$,
an inextendible causal curve $\mu$---which may be of the lines 
$\chi=\pi/2$ itself---following the proof discussed in the previous section, 
one cannot apply the remaining arguments using the sequence of
timelike geodesics $\mu_n$ to lead a contradiction, as $\mu$ would not 
contain a pair of conjugate points on it.

\medskip 
If we allow the AdS conformal boundary $\partial M$ to be dynamical and/or 
satisfy the generic condition in $\widehat M$ 
so that $\chi = \pi/2$ lines or $\mu$ can contain a pair of conjugate points, 
then the double covering approach may work. 
In the case, however, the asymptotic symmetries in the original spacetime 
might possibly break down. 
[See e.g. \cite{HM04}, for various possible boundary conditions 
in asymptotically AdS spacetimes.] 
It would be interesting to clarify possible dynamical boundary conditions 
on AdS conformal boundary that are generic enough so as to allow 
for the occurrence of a pair of conjugate points on every 
timelike geodesic in AdS conformal boundary, and at the same time,
stringent enough so as to allow meaningful notion of asymptotic
structure of AdS spacetimes. 
 
\section{A singularity theorem in spherically symmetric systems}
\label{sec:sss}

In the previous section~\ref{subsec:diff} we have seen difficulties in 
removing condition for strong gravity; without imposing the condition~(c), 
the global method used in standard proof of the singularity 
theorems does not appear to work as we have to then deal 
with a non-globally hyperbolic region including part of AdS 
conformal infinity, in which the argument about the existence 
of maximum length causal curves does not apply. 
In this section we take a different approach to singularity theorems: 
we restrict our attention to a specific class of asymptotically AdS 
spacetimes and matter field. 
Since we have already seen that, once a trapped set is formed, then
Hawking and Penrose's theorem applies, the main purpose here is to 
show the occurrence of a trapped set in asymptotically AdS spacetimes. 

\medskip 
The numerical results of Bizon and Rostworowski~\cite{Bizon} may be 
physically interpreted as follows. 
The dispersion of the scalar field increases during the time evolution and the amplitude tends to decrease when the wave packet travels 
toward the AdS boundary. However, near the AdS boundary, the effective 
potential makes the dispersion smaller by the reflection. 
Then, after several reflection, the wave packet can become supercritical 
near the center of spherical symmetry and finally collapses to a black hole.

\medskip 
Some basic aspects of gravitational collapse can be described by 
the Raychaudhuri equation for a congruence of non-spacelike curves. 
For example, let us consider an asymptotically AdS gravity system composed of 
a dust fluid with energy density $\mu>0$ and a negative cosmological constant 
$\Lambda<0$. Then, a congruence of timelike geodesics obeys the Raychaudhuri 
equation~(\ref{eqn:Raychaudhuri}) with  
$R_{\mu \nu}K^\mu K^\nu = 4\pi\mu + |\Lambda|$. 
%
%
Since all the terms in the right-hand side are non-positive, 
the expansion $\theta$ becomes larger and larger in negative 
along the geodesics, and nothing stops the congruence getting smaller and 
smaller toward a conjugate point. This means that as the flow lines of 
the dust are timelike geodesics that also obey (\ref{eqn:Raychaudhuri}), 
the energy density blows up unboundedly toward the conjugate point, 
and a singularity eventually forms.  
It is clear from Eq.~(\ref{eqn:Raychaudhuri}) that the negative 
cosmological constant plays a role to enhance the gravitational contraction.

\medskip 
This simple argument, however, does not immediately apply to the case 
with other matter models such as massless scalar fields or perfect fluids 
since the energy flow lines or the histories of small particles with non-zero 
pressure are not geodesic; they are accelerated. 
To be more precise, consider a 
perfect fluid in asymptotically AdS spacetime. 
Let $\mu$ and $p$ be the energy density and pressure of 
the fluid and let $V^\mu$ be the unit tangent vector field along 
the flow lines~($g({\bm V}, {\bm V})=-1$). 
Then, the Raychaudhuri equation of the congruence becomes 
\begin{align}
\label{Raychaudhuri-eq:p-fluid} 
\frac{d\theta}{ds}= - 4\pi(\mu + 3p) -2\sigma^2 
-\frac{1}{3}\theta^2 - |\Lambda| + \nabla_\mu\dot{V}^\mu \,.     
\end{align}
The first term is non-negative under the strong energy condition. 
The last term describing the divergence of the acceleration, 
${\dot{V}^\mu}:=V^\nu\nabla_\nu V^\mu$, of the flow lines is related to the pressure gradient, 
as seen in Eq.~(\ref{eq-perfectfluid2}). In particular, the last term 
$\nabla_\mu\dot{V}^\mu$ is not necessarily negative and can represent 
repulsion between the fluid lines when it is positive. 
Therefore it is, in principle, possible that the last term and the others 
are balanced to realize an equilibrium, stationary configuration. 
%
It may be possible that even if the right-hand side is not always balanced 
and the expansion is negative at some initial time, it could change 
its sign to positive later on and the fluid lines would bounce off 
near the center and evolve outward. 
In this case, either the fluid bounces off at AdS boundary and 
recollapses to form singularities, or the system could be dynamical 
but admit a discrete time symmetry, i.e., it repeats expansion 
and contraction cyclically without ever forming singularities.

\medskip 
In the following we will show that for a certain class of matter 
fields, a singularity must inevitably form in spherically symmetric, 
asymptotically AdS spacetime. 
A key element of our argument is that as discussed above, 
the negative cosmological constant $\Lambda$ plays a role to effectively 
confine the system in an finite region characterized by its curvature radius. 
This, in turn, implies that the magnitude of the acceleration of the flow 
lines of matter fields should be bounded from above in certain sense 
so that the flow lines can never reach AdS infinity but instead must always 
reflect back inward the spacetime.\footnote{ 
For this we need to control the term $\nabla_\mu\dot V^\mu$ 
in eq.~(\ref{Raychaudhuri-eq:p-fluid}). 
If we assume, for instance, the local condition, 
$\nabla_\mu\dot V^\mu<|\Lambda|$, then we are immediately done. 
To show our singularity theorem we shall adopt a much weaker condition, 
(\ref{condi:averaged-convergence}), than this. 
}
Then, as can also be observed in Bizon and Rostworowski's massless 
scalar field system, we expect that even if the expansion of the matter flow 
lines can change its sign and gets positive, its time average over 
sufficiently large time interval (at least larger than the time scale 
specified by the cosmological constant) should be negative. If this is 
the case, then even though the flow lines of matter fields may have to 
bounce back and forth between the center of the spacetime and AdS-boundary 
repeatedly, 
a large part of the flow lines will eventually enter a compact region near the 
center with sufficiently small radius (compatible with the hoop conjecture) 
and form a closed trapped surface.  

\bigskip
From different point of view, it is expected that such a singularity 
formation in asymptotically AdS spacetime occurs when the matter system 
under consideration admits no equilibrium configuration. 
The massless scalar field, for example, has no equilibrium configuration. 
On the other hand, charged fluid has an equilibrium configuration known 
as ``electron stars''~\cite{HartnollTavanfar}. 
In the gravity system coupled to the charged fluid, dynamical instability will not 
occur, and hence the equilibrium configuration can be realized after the time 
evolution for any regular initial data except a particular 
case~\footnote{For example, if initial spacelike 
hypersurface possesses a trapped surface, the time evolution results in the formation 
of singularity.}. 
So, we will need to impose a kind of dynamical instability condition 
that ensures no equilibrium state to appear as an end state of the dynamical 
evolution.

\bigskip 
In what follows we shall focus on spherically symmetric spacetimes with a 
negative cosmological constant $\Lambda<0$ and a perfect fluid 
with $\mu$ and $p$ being, respectively, the energy density and 
the pressure in four-dimensions.\footnote{
Our argument does not work for the vacuum case with only a cosmological 
constant. 
} 
We assume that there is a regular, initial partial 
Cauchy surface whose topology is ${\bf R}^3$ as in the standard 
global AdS spacetime. 
A generalization of our arguments below to higher dimensions should be 
straightforward. We assume that our spaceitmes are asymptotically, 
globally anti-de Sitter, and the asymptotic fall off conditions for 
the perfect fluid are defined accordingly.\footnote{ 
If a spherically symmetric configuration of the fluid has compact support 
on the initial partial Cauchy surface, then the outside of the support must 
be locally isometric to the Schwarzschild-AdS metric 
due to the Birkhoff's theorem, and therefore the event horizon and 
the central singularity must occur, unless a static configuration 
is realized. 
} 
These asymptotic conditions can be given in the conformal framework 
[See e.g., \cite{HIM05}]. 
For later convenience, we shall adopt the following coordinate system   
\begin{align}
\label{metric} 
ds^2=-f(t,r)dt^2+ h(t,r) dr^2 + R^2(t,r)d\Omega^2, 
\end{align}
where the time coordinate $t$ is taken along the fluid flow lines labelled 
by $r$, so that the unit future-directed tangent vector field 
is given by $V^\mu=f^{-1/2}(\p/\p t)^\mu$. 
Outside the support of the fluid, we assume that the time coordinate $t$ 
extends smoothly toward the AdS infinity at $r\rightarrow \infty$. 
Near the center $r=0$, due to the existence of the pressure, the timelike 
congruences of the fluid lines would never encounter a conjugate point, 
so that the coordinate system would not break down, as far as no singularities 
form. Assuming that $V^\mu$ is everywhere timelike and its orbits are 
complete, we shall show that a trapped region must form. 

\medskip  
The perfect fluid under consideration obeys  
\begin{subequations}
\begin{align}
& V^\nu \nabla_\nu \mu+(\mu+p)\nabla_\nu V^\nu=0 \,, 
\label{eq-perfectfluid1} \\
& (\mu+p)\dot{V}^\mu+(g^{\mu\nu}+V^\mu V^\nu)\nabla_\nu p=0 \,. 
\label{eq-perfectfluid2}
\end{align}
\end{subequations} 
If we define $\rho$ as a number density of the fluid particles, 
$\rho$ satisfies 
\begin{align}
\label{def-epsilon}
\mu=\rho(1+\eta), \qquad p=\rho^2\frac{d\eta}{d\rho}, 
\end{align}
where $\eta$ is the internal energy. Then, substitution of Eq.~(\ref{perfectfluid-relation}) into 
these two equations yields 
\begin{align}
\label{internal_energy}
\eta=A\rho^{\gamma-1}-1, \qquad \mu=A\rho^\gamma, 
\end{align}
where $A$ is a positive constant. 
The number current $j^\mu$ is also defined 
as $j^\mu:=\rho V^\mu$. So, we have a conservation law: 
\begin{align}
\label{conservation}
\nabla_\mu j^\mu=0. 
\end{align}

\medskip 
Since our main interest now on is the formation of a closed trapped surface 
in a spherically symmetric spacetime and thus concerns the behavior of 
the expansion $\theta$ of the fluid lines in the vicinity of the center,  
we shall consider below conditions that are required to hold for 
the fluid lines labelled by $0\leq r<r_0$ with some positive 
(possibly sufficiently small) value $r_0$. 
%
At the same time, we are interested in situations in which the flow lines 
bounce off many times near the center by its pressure gradient and 
somewhere away from the center by the AdS curvature. 
Thus, we consider cases in which expansion of the fluid lines labelled by 
$r(<r_0)$ change many times its sign, minus for contraction and plus for 
expansion after bouncing off at the center. The first condition concerns 
geometric properties and the next two (ii), (iii) the dynamics, 
and (iv) the energy condition of the fluid.

\medskip 
\noindent 
{\bf Condition (i):} \\
Let $X(t,r)$ collectively denote all physically relevant (gauge invariant) 
quantities such as $\theta$, $\nabla_\mu\dot{V}^\mu$, and $\mu$, and 
suppose $X$ be a smooth function of $r$ in the ball $0\leq r < r_*$ 
with some relevant radius $r_*$. 
%
%
(a) Suppose there is a set of $t=$constant hypersurfaces on which 
$|X(t,0)|\ge K_0$ holds for any fixed positive number $K_0$. 
We require that there is a positive constant $r_0(<r_*)$, independent of $t$, 
such that for any $r$ in the range $0\le r \le r_0$,  
\begin{align}
|X(t,r)-X(t,0)|<\frac{1}{2}|X(t,0)| \,. 
\end{align} 
%
%
%
(b) If, instead, $X(t,r_i)=0$ at $t=t_i'$ for $r=r_i~(0\le r_i\le r_0)$, 
then we require that there is a positive constant $K$~(independent of $t_i'$) 
such that for $\forall r ; \: 0\le r\le r_0$,   
\begin{align}
|X(t_i',r)|< K \,. 
\end{align}

\bigskip 
As in condition~(i), the expansion $\theta$ of $V^\mu$ is a smooth function 
of $t$ and $r$ in $0\le r\le r_0$. 
Now one can view $\theta$ as a smooth function of proper time $s$ 
with fixed label $r(<r_0)$.  
Let $s_1$ be the proper time of the flow 
lines at which $\theta$ takes a local minimum (if exists) with $\theta(s_1)<0$ 
for the first time in the future of the initial surface, and similarly 
$s_2$ be such a time at which $\theta$ takes a local minimum 
with $\theta(s_2)<0$ for the first time after $s_1$, and so on. 
Then the following two cases can occur. Case~(A): $\theta$ has only a finite 
number of local minima, or even does not admit a local minima 
from the beginning. 
Case~(B): Such a local minimum of $\theta$ occurs infinitely many times. 
We require 

\begin{figure}[h]\begin{center}
 \includegraphics[width=7cm]{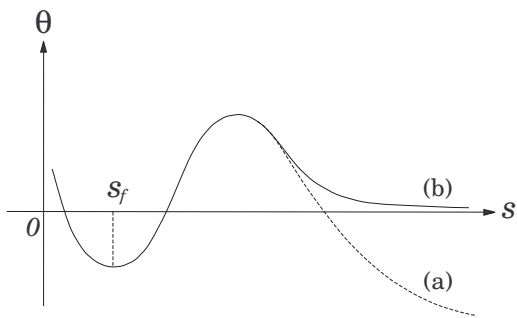} \hspace{1cm} 
 \caption{
 \label{b}
 \small{Two typical cases regarding the condition (ii) are shown 
in Case~(A): $s_f$ is the time when $\theta $ has its local minimum last.  
The condition (ii) is satisfied in the branch (a), while it is violated in 
the branch (b). 
       } }
 \end{center}\end{figure}

\medskip 
\noindent 
{\bf Condition~(ii):} \\
In Case~(A),  
there exist a time $t_0>0$ and a positive constant $c'$ such that 
$\theta(t)\le-c'$ for $\forall t \ge t_0$ and $\forall r$ in $0\le r\le r_0$ 
(i.e., $\theta$ does not increase if the minimum does not occur. 
See figure \ref{b}). 
In Case~(B), there exists a positive constant $c$ 
such that for $\forall r$ in $0< r< r_0$ the following inequalities hold;   
\begin{align}
\int^{s_{i+1}}_{s_i}\left(4\pi(\mu + 3p)+2\sigma^2 
+\frac{1}{3}\theta^2 + |\Lambda| - \nabla_\mu \dot{V}^\mu\right)ds\ge c 
\,.   
\label{condi:averaged-convergence}
\end{align} 

\medskip 
Physically this condition implies that the gravitational contraction effect 
is stronger than the effect of the repulsion by pressure gradient 
in a sense of ``time average.'' 
In what follows we call this condition the {\em averaged convergence 
condition}. 
For the case in which this condition is not satisfied, one may be able to 
find an equilibrium, static 
configuration of matter field in asymptotically AdS 
spacetimes. In Appendix, we discuss such a static configuration 
cannot be realized for perfect fluid satisfying certain equation of state. 

\medskip
\noindent 
{\bf Condition~(iii):} \\
Given a hypersurface $t=t(r)$, consider, on $t=t(r)$, the proper radial 
length $l(r_0)$ from the center to $r_0$ and the area radius $R(t,r_0)$ 
of the spherical shell labelled by $r_0$. 
We require that $l(r_0)$ and $R(t,r_0)$ be bounded above by some positive 
constants;  
\begin{align}
R(t,r_0)\le R_0 \,, \qquad 
l(r_0) := \int^{r_0}_0 
          \sqrt{
                h^2(t(\tilde{r}),\tilde{r})
              - f (t(\tilde{r}),\tilde{r})t'^2(\tilde{r})
               }d \tilde{r} 
       \le l_0\,,    
\end{align}
where $R_0$ and $l_0$ denote some time-independent positive constants. 
%

%
\medskip 
This condition comes from the role of the AdS boundary that 
confines the system in a finite region. 

\medskip 
\noindent 
{\bf Condition~(iv):} \\  
The equation of state of perfect fluid satisfies the following
; $\mu >0 $ and 
\begin{align}
\label{perfectfluid-relation}
  p=(\gamma-1)\mu, \qquad 1 < \gamma<2 \,. 
\end{align}

%
\medskip 
Now we wish to establish the following theorem: \\

\bigskip 
\noindent 
{\bf Theorem.} \\
{\em If the conditions (i)-(iv) are satisfied, 
then a future trapped surface necessarily forms in the future of 
the initial surface.}

\medskip 
Once the occurrence of a closed trapped surface, ${\cal S}$, is shown
\footnote{
%
It is worth mentioning the numerical work~\cite{harada} showing that 
a (globally) naked singularity appears in the perfect fluid when 
$1<\gamma\le 1.0105$ in asymptotically flat case.  
At first look, even though the calculation of ~\cite{harada} does not 
include cosmological constant, our lemma may appear to contradict 
the result of~\cite{harada} because we show the existence of a trapped 
surface, indicating that the resulting singularity is enclosed by the event 
horizon. However, we are assuming here that the timelike congruences near 
the center are future complete (The proper time is supposed to be extendible 
at least until that far future that a closed trapped surface is formed). 
So, this indicates that for some initial 
data with $\gamma\le 1.0105$, a timelike congruence is incomplete and 
a central singularity inevitably appears before a trapped surface 
would form. 
}
, the Hawking-Penrose theorem discussed in section~\ref{sec:basics} 
can be applied. We have 

\bigskip 
\noindent 
{\bf Corollary.} \\
{\em If the conditions~(i)-(iv) are satisfied, then any spherically 
symmetric, asymptotically AdS spacetime with perfect fluid is 
not timelike or null geodesically complete. 
}  


\medskip 
Note that the case $\gamma=1$ in condition~(\ref{perfectfluid-relation}) 
corresponds to a spherically symmetric dust fluid in asymptotically AdS 
spacetime. The dynamics of such a spacetime is described by the Lemaitre-Tolman-Bondi metric with cosmological constant~$\Lambda$.   
In this case, the last term in the Raychaudhuri 
equation~(\ref{Raychaudhuri-eq:p-fluid}) disappears. 
In this case it has been shown that 
either black holes or naked singularities form as the end state of 
the dust collapse. 
[see e.g. \cite{DJCJ} and references therein]. 
Therefore in the following we focus on the $ 1< \gamma <2$ case.


\bigskip 
\noindent 
{\it Proof.}\,) 
We first note that under the required conditions, in particular, 
condition~(ii), we have either one of the following cases.   
Case~(A): the expansion $\theta$ of the tangent vector $V^\mu$---viewed as 
a function of $t$ with fixed $r(<r_0)$---admits local minima only a finite 
number of times and afterward behaves monotonically with $t$,  
or the local minimum never occurs 
in the future of initial surface. 
%
Case~(B): the local minima occur infinite number of times (on the
assumption, of course, of geodesic completeness, that is, the spacetime 
under consideration could be extended as far future as one wishes). 


\medskip 
We consider these two cases separately. 
Let us start with Case~(A) that a local minimum disappears within finite 
times. By condition~(ii), $\theta(t,r)<-c<0$ for $t\ge t_0$ and 
$0\le r\le r_0$. Define a volume $V(t,r_0)$ enclosed by $r=r_0$ 
ball at $t=\mbox{const.}$ spacelike hypersurface: 
\begin{align}
V(t,r_0):=\int^{r_0}_0 v(t,r)drd\Omega,  
\end{align}
where $v=\sqrt{h}R^2$ and the expansion $\theta$ is rewritten by $v$ as
\begin{align}
\theta=\frac{1}{v} \frac{dv}{ds} \,. 
\end{align}
Since $\theta(t,r)<-c$ for $0\le \forall r\le r_0$, 
the volume element $v$ at any $r \le r_0$ approaches zero, 
implying $V\to 0$ at $s\to \infty$. Now take a maximal surface $\Sigma$ 
(that is, a surface whose extrinsic curvature has the vanishing trace) 
in the future of $t=t_*~(t_*\to \infty)$ const. hypersurface. 
The unit normal future-directed vector $\xi$ on ${\Sigma}$ and 
unit outward pointing vector $\eta$ normal to 
two-sphere of fixed radius in $\Sigma$ are written by 
\begin{align}
\xi^\mu= V^\mu \cosh \alpha + n^\mu \sinh \alpha \,, \qquad 
\eta^\mu= V^\mu \sinh \alpha + n^\mu \cosh \alpha \,, 
\end{align}
where the boost parameter $\alpha$ is taken to be positive (so that 
$\sinh \alpha > 0$) and where $n$ is a unit normal outward vector 
on $t=\mbox{const.}$ hypersurface, i.~e.~, $g(V,n)=0$, $g(n,n)=1$. 
Let us define the energy density $\hat{\mu}$ and the current $\hat{\pi}$ on 
$\Sigma$ as 
\begin{align}
\hat{\mu}:=T_{\mu\nu}\xi^\mu\xi^\nu-\frac{\Lambda}{8\pi} \,, \qquad 
\hat{\pi}=-\frac{1}{8\pi}R_{\mu\nu}\xi^\mu \eta^\nu
=-T_{\mu\nu}\xi^\mu \eta^\nu \,.   
\end{align}
Note that here we define $T_{\mu\nu}$ so as to 
include a negative cosmological constant term $-\Lambda g_{\mu\nu}/8\pi$ 
and therefore $T_{\mu\nu}$ above is slightly different from 
that in (\ref{condi:timelike:convergence}).
%
%
Then, $\hat{\mu}-\hat{\pi}$ is calculated as 
\begin{align}
\label{dominant-energy-con}
\hat{\mu}(t(r), r)-\hat{\pi}(t(r), r)
=\{\sinh \alpha (\sinh \alpha +\cosh \alpha ) \gamma+1\}\mu(t(r), r) \,, 
\end{align}
where $t=t(r)~(>t_*)$ on ${\Sigma}$. 

\medskip 
By the conservation law~(\ref{conservation}), we have 
\begin{align}
&{\cal N}(r_0)=\int_\Omega \int^{r_0}_0 v(t_*,r)\rho(t,r) drd\Omega
\nonumber \\
&=\int_\Omega \int^{r_0}_0 
  \rho(t(r),r)\cosh \alpha(t(r),r) d\hat{V}\Biggl{|}_{{\Sigma}}
\nonumber \\
&=\int_\Omega \int^{r_0}_0 \rho(t(r),r) v(t(r),r)drd\Omega \,,  
\end{align}
where $d\hat{V}$ is the volume element on $\Sigma$.  
Since $v(t(r),r)$ goes to zero for $0\le r\le r_0$ in the limit 
$t_*\to \infty$, 
$\rho(t(r),r)$ diverges in the limit. Owing to the relation 
$\mu=A\rho^\gamma$~(\ref{internal_energy}), 
$\mu(t(r),r)$ also diverges in the limit.  

\medskip 
Following Ref.~\cite{guven}, we shall define the total material energy $E$ 
and the radial material momentum $P$ enclosed by $r=r_0$ shell on 
the hypersurface $\Sigma$ as 
\begin{align}
\label{def-energy}
(E,\,P):=\int_\Omega \int^{r_0}_0 
         \left( 
               \hat{\mu}(t(r),r)+\frac{\Lambda}{8\pi}, \,\hat{\pi}(t(r), r) 
         \right) d\hat{V}
\end{align}
Then, by Eq.~(\ref{dominant-energy-con}), we have 
\begin{align}
 E-P&=\int_\Omega \int^{r_0}_0
      \left\{ 
             \hat{\mu}(t(r),r)+ \frac{\Lambda}{8\pi}-\hat{\pi}(t(r), r) 
      \right\}d\hat{V} 
\nonumber \\
&= \int_\Omega \int^{r_0}_0 
   \left\{
         \left(
         \sinh \alpha (\sinh \alpha + \cosh \alpha) \gamma + 1
         \right) 
         \mu(t(r), r)
         + \frac{\Lambda}{8\pi} 
   \right\}d\hat{V} 
\nonumber \\
&=4\pi \int^{r_0}_0 
   \left\{
         \frac{\left(
                 \sinh \alpha (\sinh \alpha + \cosh \alpha)\gamma+1 
               \right) 
              }{\cosh \alpha} 
        \mu(t(r), r)+ \frac{\Lambda}{8\pi \cosh \alpha}
   \right\} 
   v(t(r),r)dr 
\nonumber \\
& \ge 4 \pi \int^{r_0}_0
        \left( 
              \mu(t(r), r)+ \frac{\Lambda}{8\pi} 
        \right) v(t(r),r)dr 
\nonumber \\
& = 4\pi \int^{r_0}_0 
       \left(
             A\rho^\gamma(t(r), r) + \frac{\Lambda}{8\pi} 
       \right)
       v(t(r),r)dr
\nonumber \\ 
& > 4\pi \int^{r_0}_0 
     \left( 
           A\rho(t(r), r) + \frac{\Lambda}{8\pi} 
     \right) v(t(r),r)dr 
\nonumber \\ 
& \ge A{\cal N}(r_0)+\frac{\Lambda}{8\pi}V(t_*,r_0) \,. 
\end{align}
Here, we have used the fact that 
\begin{align}
        \sinh \alpha (\sinh \alpha+ \cosh \alpha)\gamma+1 
 \ge \cosh \alpha \,, \qquad  
\rho^\gamma(t(r), r)>\rho(t(r), r), \, \qquad v(t(r),r)<v(t_*,r)\,.   
\end{align}

\medskip 
Since $V(t_*,r_0)\to 0$ for $t_*\to \infty$ and $\rho(t(r),r)\to \infty$ 
for $0\le \forall r\le r_0$ when $t_*\to \infty$ and $\gamma>1$, 
the dominant energy condition should be satisfied for $0\le r\le r_0$ 
and $E-P$ should also diverge when $t_*\to \infty$. 
By the condition (iii), the proper radial length $l(r_0)$ along the maximal 
surface is bounded from above, implying 
\begin{align}
E-P\ge l(r_0)
\end{align}
when $t_*\to \infty$. According to Ref.~\cite{guven}, this yields 
a future apparent horizon, due to the condensation of positive energy enclosed 
by the small ball $0\le r\le r_0$ on the maximum surface. 
Thus, Case (A) follows.  
\\

\medskip 
We turn to Case~(B), i.e., when the minimum occurs infinite times. 
Integrating the Raychaudhuri equation
\begin{align}
\label{Raychaudhuri-eq} 
\frac{d\theta}{ds}= -R_{\mu\nu}V^\mu V^\nu-2\sigma^2
-\frac{1}{3}\theta^2+\nabla_\mu\dot{V}^\mu \,,  
\end{align}
we immediately obtain 
\begin{align}
\theta_{s_{i+1}}-\theta_{s_i}<-c \, 
\end{align}
from the condition (ii). Then, repeating this $N'$ times, we finally obtain 
\begin{align}
\theta_{s_{N'}}<-N'c+\theta_{s_0} \,. 
\end{align}  
From this, it follows that for any positive integer $N$, 
one can always find $t_i$ such that 
\begin{align}
 \label{lemma}
  \theta (t_i)<-Nc \,, 
 \end{align}
for $0\le r\le r_0$.  \\

\bigskip 
Note that (\ref{lemma}) implies that the amplitude of the expansion 
(at least) grows linearly in time. 
This is in accord with the appearance of secular terms 
in non-linear perturbations of the gravity system coupled to a massless scalar 
field~\cite{Bizon} or the pure gravity system in non-spherically 
symmetric asymptotically AdS spacetime~\cite{Dias}.

\medskip 
Next, integrating Eq.~(\ref{Raychaudhuri-eq}) for a small spatial interval 
$[r, r+\Delta r]$ and a small time interval 
$[t_N, t_N+\Delta t]$, we obtain 
\begin{align}
\label{volume-expansion}
4\pi \left(\theta(r){\sqrt{h}}{R}{\Biggl |}_{s_N(r)+\Delta s(r)} 
     -\theta(r){\sqrt{h}} {R} {\Biggl |}_{s_N(r)}\right)\Delta r
&= -4\pi \int^{s_N(r)+\Delta s(r)}_{s_N(r)}\sqrt{h}R 
         \left(R_{\mu\nu}V^\mu V^\nu+2\sigma^2+\frac{1}{3}\theta^2 \right)
         ds\Delta r 
\nonumber \\
& \, 
+4\pi\int^{s_N(r)+\Delta s(r)}_{s_N(r)}R \dot{V}^\mu n_\mu ds{\Biggl |}_{r+\Delta r}
-4\pi\int^{s_N(r)+\Delta s(r)}_{s_N(r)}R \dot{V}^\mu n_\mu ds{\Biggl |}_{r} \,,   
%
\end{align}
where $t=t_N$ is the time when the expansion $\theta$ at the center takes minimum 
$\Theta_N$, and $s_N(r)$, $\Delta s(r)$ are the proper time at $t=t_N$, the proper 
time interval between $[t_N, t_N+\Delta t]$ of each $r=$const. timelike curve, 
respectively. 
To derive this equation, we have used $V_\mu \dot{V}^\mu=0$.

\medskip 
Since $\theta$ at the center takes minimum $\Theta_N$ at $t=t_N$, $\theta$ begins to 
increase from $t=t_N$. This implies, by (\ref{lemma}), that 
\begin{align}
\dot{V}^\mu n_\mu {\Biggl |}_{\Delta r, t=t_N} 
-\dot{V}^\mu n_\mu {\Biggl |}_{r=0, t=t_N} 
        >\frac{(Nc)^2}{3}\sqrt{h} \Delta r \,,  
\end{align}
By the condition~(i), one also obtains 
\begin{align}
\label{inequality-acc}
\dot{V}^\mu n_\mu {\Biggl |}_{r+\Delta r, t_N}
-\dot{V}^\mu n_\mu {\Biggl |}_{r, t_N} 
  > \frac{\epsilon(Nc)^2}{3} \sqrt{h}\Delta r {\Biggl |}_{r, t_N} 
\end{align}
for any $r$ in the range $0\le r\le r_0$, where $\epsilon$ is 
a positive value $1/2<\epsilon<1$. 

\medskip 
Now consider $t=t_N$ spacelike hypersurface. Using a spatial proper radius $l$ 
on the hypersurface, 
Eq.~(\ref{eq-perfectfluid2}) is rewritten as 
\begin{align}
(\mu+p)\dot{V}^\mu n_\mu=-p_{,l} \,. 
\end{align}
Substituting Eq.~(\ref{perfectfluid-relation}) into this equation, we obtain 
\begin{align}
\label{acceleration}
\dot{V}^\mu n_\mu=-\frac{\gamma-1}{\gamma \mu}\mu_{,l} \,. 
\end{align}
Then, the inequality~(\ref{inequality-acc}) is reduced to 
\begin{align}
-\frac{(\gamma-1)\mu_{,l}}{\gamma \mu}\Biggl|_{l+\Delta l}
+\frac{(\gamma-1)\mu_{,l}}{\gamma \mu}\Biggl|_{l}>\frac{\epsilon(Nc)^2}{3}\Delta l
\end{align} 
for $0\le l\le l_0$, where $l_0=l(r_0)$. Taking the limit $\Delta l\to 0$, we finally 
obtain 
\begin{align}
\label{inequality1}
\left(\frac{(\gamma-1)\mu_{,l}}{\gamma \mu} \right)_{,l} 
     < -\frac{\epsilon(Nc)^2}{3}\,. 
\end{align}
From the regularity at the center $r=0$, $\mu_{,l}$ must be zero at the center. So, integrating 
the inequality~(\ref{inequality1}) from $l=0$ to $l$, one gets 
\begin{align}
\label{inequality2}
\frac{\mu_{,l}}{\mu}
 \le - \frac{Nl}{l_d^2} \,, 
\end{align}
where $l_d := \sqrt{3(\gamma-1) / \epsilon \gamma c^2 N }$. 
Further integration yields 
\begin{align}
\label{inequality3}
 \mu(l)\le 
 \mu(0) 
 \exp\left[-\left(N/{2l_d^2}\right){l^2} \right]
 \,.   
\end{align} 

\medskip
Since we are considering Case~(B), 
we can take an arbitrary large integer $N$, and  
therefore the above inequality means that  
the dispersion gets small indefinitely, as $t_N$, hence $l_d^{-1}$, gets
large.  
%
%
As seen in Eq.~(\ref{inequality3}), most of the energy exists inside 
$l<l_d$ for large $N$.
Since $\rho(l_d) \ll\rho(0)$, we can conclude that $l(r=r_0)$ should be 
inside $l_d$ by the condition~(i). 
%
As $t_N \rightarrow \infty$, $l_d \rightarrow 0$, and $l(r_0)$ also goes to zero. 
Thus, it is clear that the proper spatial volume $V(t_N,r_0)$ enclosed by $r=r_0$ on $t=t_N$ spacelike 
hypersurface also goes to zero by the condition~(iii) because    
\begin{align}
 V(t_N,r_0) = 4\pi\int^{l(r_0)}_0R^2(l)dl 
            \le 4\pi\int^{l(r_0)}_0R_0^2\,dl 
            = 4\pi R_0^2\,l(r_0) 
            \rightarrow 0 \,. 
\end{align}
This means that $\mu(0)=A\rho^\gamma(0)$ diverges due to the conservation law~(\ref{conservation}).

\medskip 
Having seen that $\mu$ can become arbitrarily large near the center, 
we also see that the curvature near the center becomes arbitrarily large. 
Since the congruence begins to expand between $t_N<t<t_{N+1}$ due to the
pressure,
there is a time $t'_{N}~(>t_N)$ when the 
expansion $\theta$ among the congruence between $0\le r\le r_0$ first
becomes zero.
For later purpose, let us say that $\theta(t_N)$ becomes zero first at
$r=r_i (0<r_i<r_0)$.  
Now consider the Hamiltonian constraint equation at $t=t'_{N}$ spacelike 
hypersurface. The constraint equation is 
\begin{align}
\label{constraint}
& {\cal R}=2\Lambda+16\pi \mu-({\chi^a}_a)^2+\chi^{ab}\chi_{ab} \nonumber \\
&\quad >16\pi \mu+2\Lambda-\theta^2 \,, 
\end{align}
where ${\cal R}$ is the scalar curvature of the hypersurface and $\chi_{ab}$ 
is the second fundamental form. To derive the inequality, we have used 
the fact that $\chi^{ab}\chi_{ab}$ is positive and ${\chi^a}_a=\theta$. 

\medskip 
Since $\theta$ is non-positive among the congruences between $0\le r\le r_0$ 
during $t_N\le t\le t_N'$, the volume $V(t_N',r_0)$ enclosed by $r=r_0$ on 
$t=t_N'$ spacelike hypersurface is smaller than the volume $V(t_N,r_0)$. 
Since $V(t_N,r_0)$ goes to zero, $V(t_N',r_0)$ also goes to zero when $N\to \infty$. 
Thus, we obtain $\mu(t_N',r)\ge \mu(t_N,r)$ for $r$ in $0\le r\le r_0$.  
%
Since a finite mass is concentrated in the small neighborhood of the
symmetry center, one can expect 
that a trapped surface, $\cal S$, would form, enclosing the 
mass. We will show below this is indeed the case.  

\medskip
Let us take the metric on the $t=t_N'$ spacelike hypersurface as 
\begin{align}
\label{3dim_metric}
ds^2_3=\frac{dR^2}{k(R)}+R^2d\Omega^2 \,.  
\end{align}
Then, as shown in Ref.~\cite{Hayward94}, a marginal surface for which 
the expansion $\theta_+$ of the outgoing null geodesics becomes zero appears 
only when $k(R)=0$. 
Namely, in terms of the local mass $M(R)$ defined as $k(R)=1-2M(R)/R$, the marginal 
surface appears only when 
\begin{align}
\frac{2M(R)}{R}=1 \,.  
\end{align}
Note that $M(0)=0$ by imposing regularity at the center. 

\medskip 
In the coordinates~(\ref{3dim_metric}), the scalar curvature ${\cal R}$ is calculated as 
\begin{align}
\label{3dim_scalar}
{\cal R}=\frac{4M'(R)}{R^2} \,. 
\end{align}
Substituting Eq.~(\ref{3dim_scalar}) into Eq.~(\ref{constraint}) and integrating once, 
we obtain 
\begin{align}
\label{integral_constr}
M(r)>4\pi \int^{R(r)}_0 \mu(r)R^2 dR-\int^{R(r)}_0 
\left(\frac{|\Lambda|}{2}+\frac{\theta^2}{4}\right)R^2 dR \,.   
\end{align} 
Since $\theta(t_N',r)=0$ at $r=r_i'~(0\le r_i'\le r_0)$,  
$|\theta(t_N',r)|<K$ in $0\le r\le r_0$ by the condition~(i). 
This implies that the second term in the r.h.s. of 
Eq.~(\ref{integral_constr}) is negligible 
compared with the first term, and hence $M(r)\ge 0$ for $0\le r\le r_0$.  
Thus, it is clear that $R(r_0)$ goes to zero when $V(t_N',r_0)$ goes to zero 
because
\begin{align}
V(t_N',r_0)=4\pi\int^{r_0}_0\frac{R^2}{1-\frac{2M(R)}{R}}dR 
          >\frac{4\pi }{3} R^3(r_0) \,. 
\end{align}

\medskip 
By the conservation law~(\ref{conservation}), we have 
\begin{align}
4\pi\int^{l(r_0)}_0\rho(R(l))R^2(l) dl={\cal N} \,,     
\end{align}
where $l(r_0)$ is the proper radial length from $r=0$ to $r=r_0$. 
Since $l(r_0)\le l_0$ by the condition~(iii), 
\begin{align}
\rho(R)\ge \frac{C_2}{R^2(r_0)}
\end{align}
for $0\le r\le r_0$ by the condition (ii), where $C_2$ is some positive constant.   
By the relation $\mu(R)=A\rho^\gamma(R)$, we also obtain 
\begin{align}
\label{mu_divergence}
\mu(R)\ge \frac{C_3}{R^{2\gamma}(r_0)} \,, 
\end{align}
where $C_3$ is a positive constant. Substituting Eq.~(\ref{mu_divergence}) 
into Eq.~(\ref{integral_constr}), we obtain 
\begin{align}
M( r_0)\ge \frac{4\pi C_4}{3} R^{3-2\gamma}(r_0) \,, 
\end{align}
where $C_4$ is a positive constant. Thus, we have 
\begin{align}
\lim_{N\to \infty}\frac{2M(r_0)}{R(r_0)}=\lim_{N\to \infty}
\frac{8\pi C_4}{3}R^{2(1-\gamma)}(r_0)>1 \,, 
\end{align}
where we have used the fact $R(r_0)\to 0$ when $N\to \infty$ and the condition (iv).
This means that $k(R)$ must admit zero, hence there must appear a 
trapped surface. $\Box$ \hfill 


\section{Concluding remarks}
\label{sec:conclusion}

In this paper, we have considered singularity theorems in asymptotically 
AdS spacetimes. We first briefly reviewed the basics of singularity 
theorems and checked that Hawking and Penrose's theorems, which use 
the global methods to prove, apply to asymptotically AdS spacetimes 
without major changes in its assumptions. 
Motivated from the result of \cite{Bizon} that indicates that in a spherically 
symmetric massless scalar field model, a black hole inevitably forms 
even starting from arbitrarily small initial perturbations after repeating 
the reflection at AdS infinity many times, 
we have discussed if it is possible to show a singularity theorem 
in asymptotically AdS spacetimes by taking into account the effect of 
the reflection back at AdS infinity, and instead removing the requirement 
of a strong gravity region, i.e., a trapped set, which is one of the 
essential requirements in Hawking and Penrose's
singularity theorems. We have discussed main obstacles to do 
so in section~\ref{subsec:diff}.

\medskip
Then in the second part we have focused on the system of asymptotically
AdS gravity plus a perfect fluid with spherical symmetry and have 
investigated under what conditions a closed trapped surface generically 
appears in such a system. 
As seen in section~\ref{sec:basics} once a closed trapped surface forms  
in asymptotically AdS spacetime, the existence of a singularity
can be shown under conditions used in Hawking-Penrose singularity 
theorems. We have shown that if the ``averaged convergence condition'' 
introduced in section~\ref{sec:sss} is satisfied, a trapped surface 
necessarily forms, provided that the proper time $s$ of the fluid lines 
are complete, or the spacetime can be extended as far future from initial 
data as one wants. Apart from that, there could be the case in which 
a singularity occurs without forming a closed trapped surface. 
In this case singularities that occur would be a naked one. 
Thus, we conclude that a singularity generically appears in the spherically 
symmetric asymptotically AdS spacetime for the perfect fluid system 
considered in section~\ref{sec:sss}. 

\medskip 
The ``averaged convergence condition'' implies that the expansion of 
congruence for the fluid world lines changes its sign infinitely many times 
and the absolute value at each local minimum continues to increase as 
the spacetime dynamically evolves. As shown in section~\ref{sec:sss}, 
this makes the dispersion of the energy density profile smaller and smaller 
indefinitely. It is noteworthy that imposing the ``averaged convergence 
condition'' does not necessarily mean by itself the existence of any strong 
gravity region or a high energy density region. In this sense, 
the condition is much weaker than the condition for the existence of a closed 
trapped surface, where a sufficiently large amount of mass should 
be enclosed in a small compact region.

\medskip 
It would be interesting to generalize our results here for the perfect fluid 
system to other matter field cases such as a massless or massive 
scalar field. To define the ``averaged convergence condition'' for 
the scalar field case, we need to define the notion of the energy flow lines 
of a scalar field. Kodama vector field~\cite{kodama}, which can be 
defined for any spherically symmetric spacetimes, should be a candidate 
of such a vector field because, associated with it, one can always construct 
a conserved current.

\medskip  
It would be also interesting to explore under what circumstances, 
singularities are 
avoided to form in asymptotically AdS spacetime. 
At first sight, all known numerical and analytical results~\cite{Bizon,Dias} 
combined with our results suggest that any matter fields satisfying physically 
reasonable energy conditions inevitably cause singularity formation 
in asymptotically AdS spacetime. According to the AdS/CFT correspondence, 
this phenomenon can be interpreted as a turbulent instability 
in hydromechanics, which transfers energy to higher frequency, as mentioned in 
Refs.~\cite{Bizon,Dias}. On the other hand, it is well-known in the turbulent 
phenomena that there is also an ``inverse cascade phenomenon'' which transfers 
energy to lower frequency. In the gravity side, if repulsive force is much 
stronger than attractive force for a matter field, such an inverse cascade 
phenomenon might occur since singularity formation would be avoided due to 
the large repulsive force. To answer this question, it would be worth 
investigating a charged perfect fluid in which the charge density is much 
bigger than the mass density. If the repulsive force is stronger than the attractive force, a periodic solution might be possible, as the charged fluid 
cannot escape to AdS infinity or collapse to a singularity. 
[See e.g. \cite{sos12} for a different approach to the problem.]
To pursue these possibilities, we need to investigate, by concerted efforts 
of both numerical and analytical methods, non-linear dynamics for a variety 
of matter fields in asymptotically AdS spacetimes. 
%

\section*{Acknowledgments}
We wish to thank Vitor Cardoso, Oscar Dias, Jorge Santos, Helvi Witek 
for valuable discussions. 
We would especially like to thank Gary T.~Horowitz for valuable discussions,  
in particular, on the non-existence of maximum length curve 
and the need of cut-off radius discussed in section~\ref{subsec:diff} 
as well as on the occurrence of a closed trapped surface in the last part of 
section~\ref{sec:sss}. 
We would also like to thank Tomohiro Harada for discussions on numerical 
studies of the singularity formation by perfect fluids and Tetsuya Shiromizu 
for useful comments on the first version of the paper. 
The authors thank the Yukawa Institute for Theoretical Physics at 
Kyoto University. Discussions during the YITP workshop, YITP-T-11-08 on 
``Recent advances in numerical and analytical methods for black hole dynamics''
were very useful to complete this work. 
This work was supported in part (AI) by the Grant-in-Aid for Scientific 
Research Fund of the JSPS (C)No. 22540299 and by the Barcelona Supercomputing 
Center (BSC) under Grant No. AECT-2012-2-0005 and in part (KM) by 
MEXT/JSPS KAKENHI Grant Number 23740200.

\section*{Appendix: Non-equilibrium state}

Let us check if there is a static regular asymptotically AdS solution 
in the case of the perfect fluid under consideration.  
If the system admits a static configuration, dynamically perturbed solutions 
would approach the static equilibrium configuration, 
unless the perturbation is unstable. In this case, 
the condition~(ii) would be violated, and the expansion $\theta$ approaches 
zero at future infinity~(case (b) in figure~\ref{b}).   
As shown below, however, when $\gamma>3/2$, this is not the case. 

\medskip 
As in Ref.~\cite{HE}, we shall adopt the following coordinate system 
\begin{align}
ds^2=-\frac{dt^2}{F^2(r)}+X^2(r)dr^2+r^2d\Omega^2 \,.  
\end{align}
In the case of perfect fluid with the energy-momentum tensor: 
\begin{align}
\label{EM-tensor}
T_{\mu\nu}=\mu V_\mu V_\nu+(g_{\mu\nu}+V_\mu V_\nu)p \,, 
\end{align} 
the Einstein equations are written 
\begin{subequations}
\label{Einstein-eqs}
\begin{align}
& 4\pi(\mu+p)=\frac{X'}{rX^3}-\frac{F'}{rFX^2} \,,  
\label{Einstein-eqs-1} \\
& 8\pi p=\Lambda-\frac{1}{r^2}\left(1-\frac{1}{X^2}\right)-\frac{2F'}{rFX^2} \,,
\label{Einstein-eqs-2} \\ 
& 4\pi(\mu-p)=-\Lambda+\frac{1}{r^2}+\frac{F'}{rFX^2}+\frac{1}{r^2X^3}(rX'-X)\,. 
\label{Einstein-eqs-3}
\end{align}
\end{subequations}
By eliminating $p$ from Eqs.~(\ref{Einstein-eqs-1}) and (\ref{Einstein-eqs-2}), we obtain 
\begin{align}
\label{mu-eq}
8\pi\mu=-\Lambda+\frac{1}{r^2}\left(1-\frac{1}{X^2}\right)+\frac{2X'}{rX^3} \,. 
\end{align}
Defining the mass $m(r)$ as 
\begin{align}
\label{def-X}
X^2=\left(1-\frac{2m}{r}-\frac{\Lambda}{3}r^2 \right)^{-1} \,, 
\end{align}
$m(r)$ is obtained from Eq.~(\ref{mu-eq}) as 
\begin{align}
\label{def-m}
m(r)=4\pi\int^r_0 \mu(\tilde{r})\tilde{r}^2 d\tilde{r} \,, 
\end{align}
where the regularity condition $m(0)=0$ is imposed. 

\medskip 
By the conservation law of the energy-momentum tensor~(\ref{EM-tensor}), 
we have 
\begin{align}
\label{p-evolution}
 p'=(\mu+p)\frac{F'}{F} \,. 
\end{align}
Substitution of Eqs.~(\ref{def-X}) and (\ref{p-evolution}) into Eq.~(\ref{Einstein-eqs-2}) 
yields 
\begin{align}
p'=-(\mu+p)\frac{m+\left(4\pi p-\frac{\Lambda}{3}\right)r^3}{r\left(r-2m-\frac{\Lambda}{3}r^3 \right)} \,.  
\end{align}
Using Eq.~(\ref{perfectfluid-relation}), we finally obtain the evolution 
equation for $\mu$ as 
\begin{align}
\label{mu-evolution}
(\gamma-1)\mu'=-\frac{\gamma\mu\left(m+4\pi(\gamma-1)\mu r^3-\frac{\Lambda}{3}r^3\right)}
{r\left(r-2m-\frac{\Lambda}{3}r^3\right)} \,. 
\end{align}

\medskip 
We can determine the metric functions $X$ and $F$ via Eqs.~(\ref{def-X}) and (\ref{p-evolution}) 
by solving Eqs.~(\ref{mu-evolution}) and (\ref{def-m}) by imposing the 
regularity condition at the center, $\mu'(0)=0$. 
So, for a fixed $\Lambda$ and $\gamma$ 
the only free parameter is $\mu(0)$.  
By Eq.~(\ref{mu-evolution}), we can easily check that there is no star like solution such as 
\begin{align}
\label{eq-star}
\mu>0 \qquad \mbox{for} \qquad r<R \qquad \mbox{and} \qquad 
\mu=0 \qquad \mbox{for} \qquad r\ge R \,,  
\end{align}
Suppose the existence of the solution satisfying Eq.~(\ref{eq-star}). Then, just inside of $R$, 
$r<R$, 
Eq.~(\ref{mu-evolution}) behaves as $\mu'\sim -C\mu$, where $C$ is a positive constant. 
The solution is $\mu\sim Ae^{-Cr}$ for some constant $A$, which fails 
to satisfy Eq.~(\ref{eq-star}). This indicates that the energy density $\mu$ 
decays toward the infinity but never becomes zero at any finite radius $R$.  

\medskip 
The asymptotic behavior of $\mu$ is easily obtained by Eq.~(\ref{mu-evolution}). Since $\mu$ 
decays toward the infinity, $\Lambda$ term in Eq.~(\ref{mu-evolution}) is dominant at $r\to \infty$. 
So, Eq.~(\ref{mu-evolution}) is reduced to  
\begin{align}
(\gamma-1)\mu'\simeq -\frac{\gamma \mu}{r}
\end{align}
with the asymptotic solution $\mu\sim r^{-\frac{\gamma}{\gamma-1}}$. 
If we impose a boundary condition at the infinity as $m\sim \mbox{const.}$, 
we must take $\gamma$ as 
\begin{align}
\gamma<\frac{3}{2} \,. 
\end{align}
This implies that any static configuration satisfying asymptotically AdS 
spacetime cannot be realized when  
\begin{align}
\frac{3}{2}\le \gamma<2 \,.  
\end{align}


\end{document}